\documentclass[a4paper,12pt]{article}

\usepackage{graphicx}
\usepackage{amsmath}
\usepackage{amssymb}
\usepackage{hyperref}
\usepackage{tabularx}
\newcolumntype{M}{>{$}c<{$}}
\usepackage[matrix,arrow,curve]{xy}

\numberwithin{equation}{section} \numberwithin{figure}{section}
\numberwithin{table}{section}

\def\papertitlepage{\baselineskip 3.5ex\thispagestyle{empty}}
\def\Title#1{\baselineskip 1cm \vspace{1.5cm}%
  \begin{center}{\Large\bf #1}\end{center}\vspace{0.5cm}}
\def\Authors#1{\begin{center}\renewcommand{\thefootnote}{\fnsymbol{footnote}}{\it #1}\end{center}}
\def\Abstract{\vspace{1.0cm}%
  \begin{center}{\large\bf Abstract}\end{center}}

\renewenvironment{thebibliography}{\pagebreak[3]\par\vspace{0.6em}
\begin{flushleft}{\large \bf References}\end{flushleft}
\vspace{-1.0em}

\begin{enumerate}\if@twocolumn\baselineskip=0.6em\itemsep -0.2em
\else\itemsep -0.2em\fi\labelsep 0.1em}{\end{enumerate} }

\def\del{\partial}

\DeclareMathDelimiter{\lcolon}{\mathopen}{operators}{"3A}{largesymbols}{"3A}
\DeclareMathDelimiter{\rcolon}{\mathclose}{operators}{"3A}{largesymbols}{"3A}

\def\+{\!\!+\!\!}

\def\lpar{(\!(}
\def\rpar{)\!)}

\def\dynkin(#1){(#1)}

\def\bra<#1|{\langle#1|}
\def\ket|#1>{|#1\rangle}
\def\braket<#1|#2>{\langle#1|#2\rangle}
\def\llangle{\langle\!\langle}
\def\rrangle{\rangle\!\rangle}
\def\bbra<#1|{\llangle#1|}
\def\kket|#1>{|#1\rrangle}
\def\bbraket<#1|#2>{\llangle#1|#2\rrangle}

\begin{document}
{\papertitlepage \vspace*{0cm} {\hfill
\begin{minipage}{4.2cm}
IFT-P.010/2008\par\noindent June, 2008
\end{minipage}}
\Title{Pure Spinor Partition Function Using Pad\'{e} Approximants}
\Authors{{\sc E.~Aldo~Arroyo${}$\footnote{\tt aldohep@ift.unesp.br}}
\\
${}$Instituto de F\'{i}sica Te\'{o}rica, S\~{a}o Paulo State University, \\[-2ex]
Rua Pamplona 145, S\~{a}o Paulo, SP 01405-900, Brasil
\\
${}$ }

} 

\vskip-\baselineskip
{\baselineskip .5cm \Abstract In a recent paper, the partition
function (character) of ten-dimensional pure spinor worldsheet
variables was calculated explicitly up to the fifth mass-level. In
this letter, we propose a novel application of Pad\'{e} approximants
as a tool for computing the character of pure spinors. We get
results up to the twelfth mass-level. This work is a first step
towards an explicit construction of the complete pure spinor
partition function.
 }
\newpage
\setcounter{footnote}{0}
\tableofcontents

\section{Introduction}
A few years ago, a new formalism was proposed for quantizing the
superstring in a manifestly ten-dimensional super-Poincar\'{e}
covariant manner \cite{Berkovits:2000fe}. This formalism has been
used for computing covariant multiloop amplitudes
\cite{Berkovits:2004px}, leading to new insights into perturbative
finiteness of superstring theory \cite{Berkovits:2006vc}.

This pure spinor formalism for superstrings, in the flat background,
is based on a set of ten-dimensional superspace variables $x^m$,
$\theta^\alpha$ (and its conjugate $p_\alpha$), and a set of bosonic
ghosts fields $\lambda^\alpha$ (and its conjugate $w_\alpha$)
transforming as an $SO(10)$ spinors and satisfying a pure spinor
constraint $\lambda^\alpha \gamma^m_{\alpha\beta} \lambda^\beta=0$
for $m=0$ to $9$. The need of this constraint comes from nilpotency
of a BRST operator constructed from fermionic Green-Schwarz (GS)
constraints and the pure spinor $\lambda^\alpha$. It has been shown
that the string spectrum found using the BRST cohomology coincides
with the light-cone spectrum of the GS string
\cite{Berkovits:2000nn,Berkovits:2004tw,Aisaka:2004ga}. This fact
was also recently proved by using the pure spinor superstring
partition function \cite{PS} (up to the fifth mass-level).

The information of the spectrum is encoded in the partition function
\begin{eqnarray}
Z(q,t)\equiv Tr_{\mathcal{H}}[(-)^F q^{L_0}t^{J_0}] = \sum_h
Z_h(t)q^h \, ,
\end{eqnarray}
where $J_0=\oint dz J(z)$ and $L_0=\oint dz \, z T(z)$, $J_0$ is the
zero mode of the $U(1)$ current $J(z)$, $L_0$ is the zero mode of
the Virasoro generator $T(z)$, $F$ the fermion number (which takes
account of the statistics), and $\mathcal{H}$ the corresponding
Hilbert space. The computation of the pure spinor superstring
partition function boils down to compute the partition function of
the pure spinor sector since the calculation of the matter sector
($x^m$, $\theta^\alpha$) is very simple \cite{PS,Berkovits:2005hy,
Grassi:2005jz}.

There have been many attempts for computing the partition function
of pure spinors, for instance, in \cite{Berkovits:2005hy,
Grassi:2005jz} were computed character formulas $Z_h(t)$ up to the
level $h=1$, i.e. the zero mode $Z_0(t)$ and first massive $Z_1(t)$
part of the partition function (1.1). In \cite{Grassi:2006wh,
Grassi:2007va}, the authors have tried to find a prescription which
allows to compute the partition function of curved beta-gamma
systems, but the employment of those techniques to the case of pure
spinors are still lacking. In \cite{Toymodels} it was studied
partition functions of certain subset of curved beta-gamma systems
(defined by quadratic constraints) as toy models in order to
specialize then to the case of pure spinors.

Recently, it has been presented a formal expression for the
partition function of pure spinors in terms of an infinite set of
free-field ghosts \cite{PS}
\begin{eqnarray}
Z(q,t)=\prod_{k=1}^\infty \big[ (1-t^{k})^{-N_k}
\prod_{h=1}^{\infty} (1-q^h t^{k})^{-N_k} (1-q^h t^{-k})^{-N_k}
\big] \, ,
\end{eqnarray}
where $N_k$ are the multiplicities of the ghost fields. The use of
these ghosts cames from the resolution of the pure spinor
constraint, and the necessity of infinite many of them is because
the pure spinor constraint is infinitely reducible
\cite{PS,Chesterman:2002ey}. Although it may seem difficult to
extract useful information from the formal expression (1.2), by
appealing to some regularization procedure in order to guarantee the
convergence of the infinite product over $k$, character formulas
$Z_h(t)$ were calculated up to the fifth mass-level ($h=5$). These
first five character formulas were also computed using fixed point
techniques \cite{PS}.

In this paper, we propose a novel application of Pad\'{e}
approximants \cite{pade} as a tool for computing higher-level
character formulas $Z_h(t)$ of pure spinors. We show by explicit
computations that our first five character formulas are in agreement
with the ones found in \cite{PS}, and in addition, we obtain results
up to the twelfth mass-level ($h=12$). It is left as an ambitious
challenge to find the complete pure spinor partition function. We
hope that our work is a first step towards this explicit
construction of the pure spinor partition function.

The paper is organized as follows: In section 2, a short review of
$SO(10)$ pure spinors is given. In section 3, we review some
concepts and ideas regarding to the partition function of pure
spinors. In section 4, we describe the method of Pad\'{e}
approximants for computing higher level character formulas, and we
present our results. A summary and further interesting topics are
given in section 5. Finally in appendix A, we present some details
involved in the computations of higher level character formulas.

\section{Pure spinors: a reminder}
The $SO(10)$ pure spinor $\lambda^\alpha$ is constrained to satisfy
$\lambda^\alpha \gamma^{m}_{\alpha\beta} \lambda^\beta =0$, where
$m=1$ to $10$, $\alpha, \, \beta=1$ to $16$.
$\gamma^{m}_{\alpha\beta}$ and $\gamma^{m \, \alpha\beta}$ are $16
\times 16$ symmetric matrices which are the off-diagonal blocks of
the $32\times 32$ ten-dimensional $\Gamma$-matrices and satisfy $
\gamma^{(m}_{\alpha \gamma} \gamma^{n) \, \gamma\beta}=2 \eta^{mn}
\delta_{\alpha}^{\beta} $. This implies that $\lambda^\alpha
\lambda^\beta$ can be written as
\begin{eqnarray}
\lambda^\alpha \lambda^\beta = \frac{1}{5! 2^5}
\gamma^{\alpha\beta}_{mnpqr} (\lambda^\gamma
\gamma^{mnpqr}_{\gamma\delta}\lambda^\delta)
\end{eqnarray}
where $\lambda \gamma^{m n p q r} \lambda$ defines a $5$-dimensional
complex hyperplane. This $5$-dimensional complex hyperplane is
preserved up to a phase by a $U(5)$ subgroup of $SO(10)$ rotations.
So projective pure spinors in $d =10$ Euclidean dimensions
parameterize the coset space $SO(10)/U(5)$, which implies that
$\lambda^\alpha$ has $11$ independent complex degrees of freedom
\cite{Cartan}.

The pure spinor constraint (2.1) can be solved and we can express
$\lambda^\alpha$ in terms of the $11$ independent degrees of
freedom. This was done in \cite{Berkovits:2000fe}
 by decomposing the $16$
components of $\lambda^\alpha$ into $SU(5)\times U(1)$
representations as
\begin{eqnarray}
 \lambda^+ = \gamma, \;\;\; \lambda_{[ab]}=\gamma u_{[ab]},
\;\;\; \lambda_{[abcd]}=-\frac{1}{8}\gamma u_{[ab}u_{cd]},
\end{eqnarray}
where $\gamma$ is an $SU(5)$ scalar, and $u_{[ab]}$ is an $SU(5)$
antisymmetric two-form. Using this decomposition, and by bosonizing
the $(\beta , \gamma)$ fields as $(\beta=\partial \xi e^{-\phi},
\gamma = \eta e^{\phi})$, we can write the formulas for the currents
\cite{Berkovits:2005hy}
\begin{gather}
J=-\frac{5}{2}\partial \phi -\frac{3}{2}\eta\xi, \\
N^{ab}=v^{ab}, \nonumber \\
N_a^b=-u_{ac}v^{bc}
+\delta^b_a(\frac{5}{4}\eta\xi+\frac{3}{4}\partial \phi),
\nonumber\\
N_{ab}=3 \partial u_{ab}+u_{ac}u_{bd}v^{cd}
+u_{ab}(\frac{5}{2}\eta\xi+\frac{3}{2}\partial \phi), \nonumber\\
T=\frac{1}{2}v^{ab}\partial u_{ab} - \frac{1}{2}\partial \phi
\partial \phi - \eta \partial \xi +\frac{1}{2}\partial
(\eta\xi)-4\partial(\partial \phi+\eta\xi), \nonumber
\end{gather}
where $T$ is the stress-energy tensor and the worldsheet fields
satisfy the OPE's
\begin{eqnarray}
\eta(y)\xi(z)\sim (y-z)^{-1}, \;\;\; \phi(y)\phi(z)\sim - \log(y-z),
\;\;\; v^{ab}u_{cd}\sim \delta_c^{[a}\delta_d^{b]}(y-z)^{-1}.
\end{eqnarray}

Using these parameterizations of a pure spinor, the OPE's of the
currents in (2.3) can be computed to be
\begin{gather}
N_{mn}(y)\lambda^{\alpha}(z)\sim
\frac{1}{2}\frac{1}{y-z}(\gamma_{mn}\lambda)^\alpha, \;\;\; J(y)
\lambda^\alpha(z) \sim \frac{1}{y-z}\lambda^\alpha,\\
N^{kl}(y)N^{mn}(z) \sim
-\frac{3}{(y-z)^2}(\eta^{n[k}\eta^{l]m})+\frac{1}{y-z}(\eta^{m[l}N^{k]n}-\eta^{n[l}N^{k]m}),
\nonumber \\
J(y)J(z) \sim - \frac{4}{(y-z)^2}, \;\;\; J(y)N^{mn}(z) \sim 0,
\nonumber\\
N_{mn}(y)T(z) \sim \frac{1}{(y-z)^2}N_{mn}(z), \;\;\; J(y)T(z)
\sim -\frac{8}{(y-z)^3}+\frac{1}{(y-z)^2}J(z), \nonumber\\
T(y)T(z) \sim
\frac{1}{2}\frac{22}{(y-z)^4}+\frac{2}{(y-z)^2}T(z)+\frac{1}{y-z}\partial
T. \nonumber
\end{gather}

So the conformal central charge is $22$, the ghost-number anomaly is
$-8$, the Lorentz central charge is $-3$, and the ghost-number
central charge is $-4$. One can verify the consistency of these
charges by considering the Sugawara construction of the stress
tensor
\begin{gather}
T= \frac{1}{2(k+\delta)} N_{mn} N^{mn}+\frac{1}{8} J J +
 \partial J \, ,
\end{gather}
where $k$ is the Lorentz central charge, $\delta$ is the dual
Coxeter number for $SO(10)$. Setting $k = -3$, one finds that the
$SO(10)$ current algebra contributes $-27$ to the conformal central
charge and the ghost current contributes $49$ to the conformal
central charge. So the total conformal central charge is $22$ as
expected.

\section{Partition function of pure spinors}
In this section, we review some aspects of the pure spinor partition
function and indicate the main results we are going to use. The
characters of the states (in the pure spinor Hilbert space
\cite{PS}) we wish to keep track of are their statistics, weights
$h$ (Virasoro levels), $t$-charge (measured by $J=w_\alpha
\lambda^\alpha$). Introducing formal variables $(q,t)$ for each
quantum number, we define the character as
\begin{eqnarray}
Z(q,t)&=&\text{Tr} (-1)^F q^{L_0} t^{J_0} \nonumber \\
&=& \sum_{h \geq 0} Z_h(t) q^h \\
&=& \sum_{h \geq 0 , n} \mathcal{N}_{h,n} q^h t^n \, . \nonumber
\end{eqnarray}

The trace is taken over the pure spinor Hilbert space. The quantum
numbers of the basic operators $\omega$ and $\lambda$ are:
$h(\omega,\lambda)= (1,0), \; t(\omega,\lambda) = (-1,1)$. In
\cite{PS}, it was given a prescription for computing the
$\mathcal{N}_{h,n}$ coefficients \footnote{These coefficients
represent the number of states at each level $h$ with $n$
$t$-charge, and its sign tell us whether the states are fermionic or
bosonic.}. Once these coefficients are know, it is possible to
calculate the character of pure spinors at each level $h$
\begin{eqnarray}
Z_h(t)= \sum_{ n} \mathcal{N}_{h,n} t^n \, .
\end{eqnarray}

At the lowest level, the Hilbert space is spanned by non-vanishing
polynomials of $\lambda$. Due to the pure spinor constraint,
$\lambda$'s can only appear in the ``pure spinor representations''
\begin{align}
  \lambda^{\lpar \alpha_{1}}\lambda^{\alpha_{2}}\cdots \lambda^{\alpha_{n}\rpar}
 &= [0000n]t^{n}\,,\quad(n\ge0)\,.
\end{align}
Here, we also indicated the $t$-charge of the state, and the symbol
$\lpar \alpha_{1}\alpha_{2}\cdots \alpha_{n}\rpar$ signifies the
``spinorial $\gamma$-traceless condition'', which means that the
expression is zero when any two indices $\alpha_{i}\alpha_{j}$ are
contracted using $\gamma^{\mu}_{\alpha_{i}\alpha_{j}}$.
Since the pure spinor representations have dimension
\begin{align}
\dim[0000n] &= {(n+7)(n+6)(n+5)^2(n+4)^2(n+3)^2(n+2)(n+1) \over
7\cdot6\cdot5^2\cdot4^2\cdot3^2\cdot2} \,,
\end{align}
the level zero character is easily found to be
\cite{Berkovits:2005hy,Grassi:2005jz}
  \begin{align}
   \label{eqn:z0dim}
    Z_{0}(t) &= {1-10t^2+16t^3-16t^5+10t^6-t^8 \over (1-t)^{16} } \,.
  \end{align}

As it was analyzed in \cite{PS,Chesterman:2002ey}, the pure spinor
constraint can be resolved by introducing an infinite chain of
free-field ghosts. The multiplicities $N_k$ of the ghosts can be
obtained by writing the level zero character (3.5) as
\cite{PS,Berkovits:2005hy}
\begin{eqnarray}
Z_0(t)=\prod_{k=1}^{\infty}(1-t^{k})^{-N_k} \, .
\end{eqnarray}
The fields at ghost number $k$ will have $|N_k|$ components, and
will be bosons for $N_k > 0$ and fermions for $N_k < 0$. The
multiplicities $N_k$ contain the information about the Virasoro
central charge, as well as the ghost current algebra:
\begin{eqnarray}
\frac{1}{2}c_{\text{vir}}=\sum_{k=1}^{\infty} N_k,  \;\;\;\;\;
a_{\text{ghost}}=-\sum_{k=1}^{\infty} k N_k, \;\;\;\;\;
c_{\text{ghost}}=-\sum_{k=1}^{\infty} k^2 N_k.
\end{eqnarray}
We can easily deduce from (3.5) and (3.6):
\begin{eqnarray}
N_1=16, \; N_2=-10, \; N_3=16, \; N_4=-45, \; N_5=144, \; N_6=-456,
\; N_7=1440, \, .\,.\,. \;\;\;
\end{eqnarray}

For the computations to be done in the appendix A, we will need to
know the value of the moments of the $N_k$'s, i.e. we want:
$\sum_{k=1}^{\infty}k^{s+1}N_k$. This was analyzed in
\cite{Berkovits:2005hy}. The moments of (3.8) are given by
\begin{eqnarray}
\sum_{k=1}^{\infty}k^{s+1}N_k&=&12 -
2^{s+1}-\frac{1}{\zeta(-s)}\sum_{k=1}^{\infty} k^s \big(
(-2-\sqrt{3})^k+(-2+\sqrt{3})^k  \big)   \;\;\;  \\
&=&12 -
2^{s+1}-\frac{Li_{-s}(-2-\sqrt{3})+Li_{-s}(-2+\sqrt{3})}{\zeta(-s)}
\, , \nonumber
\end{eqnarray}
where $Li_s(z)$ is the so-called polylogarithm (also known as de
Jonqui\`{e}re's function), it is a special function defined by the
sum
\begin{eqnarray}
Li_s(z)= \sum_{k=1}^{\infty} \frac{z^k}{k^s} \, . \nonumber
\end{eqnarray}

Another way to get $\sum_{k=1}^{\infty}k^{s+1}N_k$ is by considering
the general expression of the form we analyzed in (3.5) and (3.6):
\begin{eqnarray}
\prod_{k=1}^{\infty}(1-t^{k})^{-N_k}=\frac{P(t)}{Q(t)}, \nonumber
\end{eqnarray}
where $P$ and $Q$ are some polynomials. We have
\begin{eqnarray}
\sum_{k=1}^{\infty}N_k
\log(1-\text{e}^{kx})=-\log\frac{P(\text{e}^x)}{Q(\text{e}^x)}.
\nonumber
\end{eqnarray}
Since
\begin{eqnarray}
\log(1-\text{e}^x)=\log(-x) + \frac{x}{2}+\sum_{g=1}^{\infty}
\frac{B_{2g}}{2g(2g)!}x^{2g}, \nonumber
\end{eqnarray}
where $B_k$ are Bernoulli numbers, we have:
\begin{eqnarray}
\log(x)\sum_{k=1}^{\infty} N_k+\sum_{k=1}^{\infty} \log(-k) N_k +
\frac{x}{2}\sum_{k=1}^{\infty}k N_k +
\;\;\;\;\;\;\;\;\;\;\;\;\;\;\;\;\;\;\;\;\;\;\;\;\;\;\;\; \nonumber
\\ \;\;\;\;\;\;\;\;\;\;\;\;\;\;\;\;\;\;\;\;\;\;\;\;\; \sum_{g=1}^{\infty}
\frac{B_{2g}}{2g(2g)!}x^{2g}\sum_{k=1}^{\infty} k^{2g}
N_k=-\log\frac{P(\text{e}^x)}{Q(\text{e}^x)}.
\end{eqnarray}

Using (3.5) and expanding the right hand side RHS of (3.10), we
obtain the following value for the moments
\begin{gather}
\sum_{k=1}^{\infty} N_k = 11 \;, \;\;\; \sum_{k=1}^{\infty}k N_k = 8
\; , \;\;\; \sum_{k=1}^{\infty} k^2 N_k = 4 \; , \;\;\;
\sum_{k=1}^{\infty} k^4 N_k = -4 \; ,  \\
\sum_{k=1}^{\infty} k^{6} N_k = 4 \;, \;\;\; \sum_{k=1}^{\infty}k^8
N_k = \frac{68}{3} \;\; \text{ and} \;\; \sum_{k=1}^{\infty} k^{10}
N_k = -396\, . \nonumber
\end{gather}
The first three moments of $N_k$'s given in (3.11) contain the
information about the conformal central charge $c_{\text{vir}}$, the
ghost-number anomaly $a_{\text{ghost}}$, and the ghost-number
central $c_{\text{ghost}}$ (3.7). As we can check, the value of
these current central charges are in agreement with the
non-covariant calculation (2.5).

In a recent paper \cite{PS}, using the ghosts, it was constructed a
BRST operator and since this operator was given such that it carries
zero $t$-charge, the partition function of its cohomology is equal
to that of the total Hilbert space of (now unconstrained) pure
spinors and the ghosts. Therefore, the partition function of pure
spinors was formally written as
\begin{eqnarray}
Z(q,t)&=&\prod_{k=1}^\infty \big[ (1-t^{k})^{-N_k}
\prod_{h=1}^{\infty} (1-q^h t^{k})^{-N_k} (1-q^h t^{-k})^{-N_k}
\big]  \\
Z(q,t)&=&\prod_{k=1}^\infty \big[ \prod_{h=0}^{\infty} (1-q^h
t^{k})^{-N_k} \prod_{h=1}^{\infty} (1-q^h t^{-k})^{-N_k} \big] =
\sum_{h=0}^{\infty} Z_h(t) q^h. \nonumber
\end{eqnarray}
Using (3.11) and (3.12), an elementary calculation shows
\begin{eqnarray}
Z(q,t)&=& - t^{-8} Z(q,1/t), \\
Z(q,t)&=& - t^{-4} q^{2} Z(q,q/t) \, .
\end{eqnarray}
These symmetries (3.13) and (3.14) of the partition function are
referred as field-antifield and $*$-conjugation symmetry
respectively \cite{PS}. In the forthcoming sections, starting from
the formal expression of the partition function (3.12), we are going
to describe a method for computing higher level character formulas
$Z_h(t)$.

\section{Pad\'{e} approximants}
The Pad\'{e} approximation seeks to approximate the behavior of a
function by a ratio of two polynomials. This ratio is referred to as
the Pad\'{e} approximant. This approximation works nicely even for
functions containing poles, because the use of rational functions
allows them to be well-represented. Recently, the Pad\'{e}
approximation has been applied to string field theory to analyze the
tachyon condensation
\cite{Taylor:2002fy,Schnabl:2005gv,Bagchi:2008et}.

Let us now consider the general equations of the Pad\'{e}
approximation. Given some function $f(t)$, its $[M/N]$ Pad\'{e}
approximant denoted by $f^{[M/N]}(t)$ is a rational function of the
form \cite{pade}
\begin{eqnarray}
f^{[M/N]}(t)= \frac{1+\sum_{j=1}^{M} p_j t^j }{\sum_{j=0}^{N} q_j
t^j} \, ,
\end{eqnarray}
where the coefficients $p_1$, $p_2$, $\cdots$, $p_M$, $q_0$, $q_1$,
$\cdots$, $q_N$, are obtained by solving a system of $M+N+1$
algebraic equations
\begin{eqnarray}
\frac{d^n f^{[M/N]} }{dt^n} (a) = f^{(n)}(a) \, ,
\;\;\;\;\;\;\;\;\;\;\; n=0, \, 1, \, 2, \, \cdots, \, M+N \, .
\end{eqnarray}
The equations (4.2) came from equating the coefficients of $(t-a)^n$
(up to the order $n=M+N$) in the Taylor expansion of the functions
$f(t)$ and $f^{[M/N]}(t)$ around some point $t=a$ (which usually is
taken at $t=0$).

Having sketched briefly the method to approximate functions by means
of rational functions. Next, we are going to use this method for
computing higher level character formulas of pure spinors. Let us
start by written the formal expression (3.12) for the partition
function of pure spinors like the following
\begin{eqnarray}
Z(q,t)= Z_0(t)\Big[1 +  \sum_{h=1}^{\infty} f_h(t) q^h \Big]\, ,
\end{eqnarray}
where the level $h$ function $f_h(t)$ is defined by
\begin{eqnarray}
f_h(t)=\frac{1}{h!} {\del^h \over \del q^h}\tilde{Z}(q,t)\big|_{q=0}
\;\;\;\text{where}\;\;\; \tilde{Z}(q,t)=\prod_{k=1}^\infty
\prod_{h=1}^{\infty} (1-q^h t^{k})^{-N_k} (1-q^h t^{-k})^{-N_k} .
\end{eqnarray}

As we know by a previous work \cite{PS}, up to the level $h=5$,
these level $h$ functions are given by rational functions.
Therefore, this result is an indication that these level $h$
functions can be computed by means of Pad\'{e} approximants. In fact
this is the case as it is shown in the appendix A, the functions
$f_1(t)$, $f_2(t)$, $f_3(t)$, $\cdots$ can be calculated using
Pad\'{e} approximants. As our main result, we have noted that these
functions can be written like the following
\begin{eqnarray}
f_h(t)= \frac{\sum_{i=0}^{2h+6} C_{i,h} t^i}{t^{h+2}(1 + 4 t + t^2)}
\, ,
\end{eqnarray}
where the value of the coefficients $C_{i,h}$ up to the level $h=12$
are given in the following tables
\begin{align}
\begin{tabular}{|c|c|c|c|c|c|c|c|c|}
    \hline
$i$ & $C_{i,0}$ &  $C_{i,1}$    &   $C_{i,2}$  & $C_{i,3}$  &  $C_{i,4}$    &   $C_{i,5}$  &  $C_{i,6}$ & $C_{i,7}$  \\
    \hline
0   & 0 & 0 & $-$1 & $-$16 &$-$126 &  $-$672 & $-$2772 & $-$9504  \\
1   & 0 & 0 & 12 & 146 & 920&  3996 & 13440 & 37224   \\
2   & 1 & 0 & $-$67 & $-$536 & $-$2411 & $-$7616 & $-$18358 & $-$35184 \\
3   & 4 & 46 & 248 & 822 & 1852 & 3270 & 7752 & 33356  \\
4   & 1 & 40 & 319 & 1200 & 1745 & $-$ 5944 & $-$48147 & $-$179648  \\
5   & 0 & 46 & \, 628 \, & 4114 & 17000 & 48206 & 91948 & 87730  \\
6   & 0 & 0 & 319 & \, 3720 \, & 21767  & 82112 & 210717  & 326760  \\
7   & 0 & 0 & 248 & 4114 & \, 32356 \, & 162662 & 585464  & 1575690 \\
8   & 0 & 0 & $-$67 & 1200 & 21767 & \, 162552 \,  & 778424  & 2706944 \\
9   & 0 & 0 & 12 & 822 & 17000 & 162662 & \, 977032 \,   & 4215020  \\
10  & 0 & 0 & $-$1 & $-$536 & 1745 & 82112 & 778424  & \, 4454624 \,  \\
11  & 0 & 0 & 0 & 146 & 1852 & 48206 & 585464 & 4215020  \\
12  & 0 & 0 & 0 & $-$16 & $-$2411 & $-$5944  & 210717 & 2706944  \\
13  & 0 & 0 & 0 & 0 & 920 & 3270 & 91948 & 1575690  \\
14   &0& 0 & 0 & 0 & $-$126 & $-$7616 & $-$48147  & 326760  \\
15   &0& 0 & 0 & 0 & 0 & 3996 & 7752 & 87730   \\
16   &0& 0 & 0 & 0 & 0 & $-$672 & $-$18358 & $-$179648 \\
17   &0& 0 & 0 & 0 & 0 & 0 & 13440  & 33356  \\
18   &0& 0 & 0 & 0 & 0 & 0 & $-$2772 & $-$35184  \\
19   &0& 0 & 0 & 0 & 0 & 0 & 0 & 37224  \\
20   &0& 0 & 0 & 0 & 0 & 0 & 0 & $-$9504  \\
\hline
\end{tabular}
\end{align}
\begin{align}
\begin{tabular}{|c|c|c|c|c|c|}
    \hline
$i$ & $C_{i,8}$ & $C_{i,9}$ &  $C_{i,10}$    &   $C_{i,11}$  & $C_{i,12}$     \\
    \hline
0  & $-$28314 & $-$75504 & $-$184041 & $-$416416 & $-$884884   \\
1  & 87912 & 180180 & 320892 & 484770 & 565136 \\
2  & $-$55368  & $-$77968 & $-$130185 & $-$342472 & $-$1118117   \\
3  & 145512 & 513680 & 1480688 & 3596898 & 7511244   \\
4  & $-$467078 & $-$900256 & $-$1189750 & $-$468240 & 2940853   \\
5  & $-$112192 & $-$651084 & $-$1221496 & 23356 & 8349688   \\
6  & $-$14878 & $-$1971392 & $-$7447790 & $-$17913424 & $-$30692216    \\
7  & 3130008 & 3975312 & 77136 & $-$15954844 & $-$51076344   \\
8  & 7136292  & 14067968 & 18009420 & 783936 & $-$74353372   \\
9  & 13953544 & 36499868 & 75237248 & 115010006 & 94216072   \\
10 & 18453761 & 59453552 & 153557340 & 318143976 & 504911177   \\
11 & \, 21308252 \, & 82467920 & 255938464 & 651539178 & 1363603964   \\
12 & 18453761 & \, 88624848 \, & 330202624 & 999457424 & 2512598839   \\
13 & 13953544 & 82467920 & \, 366326624 \, & 1301105402 & 3825279040   \\
14 & 7136292 &59453552& 330202624 & \, 1398947880 \, & 4802902081   \\
15 & 3130008 &36499868& 255938464 & 1301105402 & \, 5216743428  \, \\
16 & $-$14878 &14067968& 153557340 & 999457424 & 4802902081  \\
17 & $-$112192 &3975312& 75237248 & 651539178 & 3825279040   \\
18 &  $-$467078 &$-$1971392& 18009420 & 318143976 & 2512598839   \\
19 & 145512 &$-$651084& 77136 & 115010006 & 1363603964   \\
20 & $-$55368 &$-$900256&  $-$7447790 & 783936 & 504911177   \\
21 & 87912 &513680& $-$1221496 & $-$15954844 & 94216072   \\
22 & $-$28314 &$-$77968& $-$1189750 & $-$17913424 & $-$74353372  \\
23 & 0 &180180& 1480688 & 23356 & $-$51076344 \\
24 & 0 &$-$75504& $-$130185 & $-$468240 & $-$30692216  \\
25 & 0 &0& 320892 & 3596898 & 8349688  \\
26 & 0 &0& $-$184041 & $-$342472 & 2940853  \\
27 & 0 &0& 0 & 484770 & 7511244 \\
28 & 0 &0& 0 & $-$416416 & $-$1118117  \\
29 & 0 &0& 0 & 0 & 565136  \\
30 & 0 &0& 0 & 0 & $-$884884 \\
\hline
\end{tabular}
\end{align}

We have defined the values of the $C_{i,0}$ coefficients such that
the level zero function is defined as $f_0(t)=1$. It is interesting
to note that the coefficients $C_{i,h}$ satisfy the following
identities
\begin{eqnarray}
C_{i,h}&=&C_{2 h+6-i,h} \, , \\
\sum_{i=0}^{h'} \phi_{h'-i} C_{i,h}&=& - \sum_{i=0}^{h} \phi_{h-i}
C_{i,h'} \, ,
\end{eqnarray}
which can be derived by using the two symmetries of the partition
function (3.13), (3.14) and verified by using the coefficients shown
in the above tables. The coefficient $\phi_{m}$ is generated by
\begin{eqnarray}
\frac{Z_0(t)}{1+4 t +t^2}=\sum_{n=0}^{\infty} \phi_{n} t^n \, ,
\end{eqnarray}
and it is given explicitly by the formula
\begin{eqnarray}
\phi_{n}= \frac{(1 + n) (2 + n) (3 + n) (4 + n) (5 + n)^2 (6 + n) (7
+ n) (8 + n) (9 + n)}{2^7 \cdot 3^4 \cdot 5^2 \cdot 7} \, .
\end{eqnarray}

The importance of the identity (4.9) is as follows. If we know the
coefficients $C_{i,0}$, $C_{i,1}$, $\cdots$, $C_{i,h'}$, it is
possible to compute explicitly the coefficients $C_{0,h}$,
$C_{1,h}$, $\cdots$, $C_{h',h}$. For instance, setting $h'=0$ in
equation (4.9), we get
\begin{eqnarray}
C_{0,h}= - \sum_{i=0}^{h} \phi_{h-i} C_{i,0} \, ,
\end{eqnarray}
using (4.11) and the value of the coefficients $C_{i,0}$ given in
the table (4.6) into the equation (4.12), we obtain
\begin{eqnarray}
C_{0,h}= \frac{(1-h) h (1 + h)^2 (2 + h)^2 (3 + h)^2 (4 + h) (5 +
h)}{2^6 \cdot 3^3 \cdot 5^2 \cdot 7} \, .
\end{eqnarray}
By employing the same steps given above, setting $h'=1$ in equation
(4.9), we arrive to the following expression for the $C_{1,h}$
coefficient
\begin{eqnarray}
C_{1,h}= \frac{(h - 1) h (1 + h)^2 (2 + h) (3 + h) (4 + h) (108 + 10
h + 12 h^2 - h^3)}{2^5 \cdot 3^3 \cdot 5 \cdot 7} \, .
\end{eqnarray}

Finally, it would be important to find an explicit expression for a
general coefficient $C_{i,h}$ (for all $h\geq0$ and $i\geq 0$). It
is clear that if we know explicitly $C_{i,h}$, it should be possible
to write a compact expression for the complete pure spinor partition
function
\begin{eqnarray}
Z(q,t)= \frac{1+t\;\;\;}{(1-t)^{11}} \sum_{h=0}^{\infty}
\sum_{i=0}^{2h+6} C_{i,h} t^{i-h-2} q^h \, ,
\end{eqnarray}
where the factor $(1+t)/(1-t)^{11}$, in front of our formula (4.15),
comes from substitution of equations (3.5) and (4.5) into the
equation (4.3).

\section{Summary and discussions}
We have given in detail a prescription for computing the partition
function of pure spinors. This prescription is mainly based in the
knowledge of the zero mode part $Z_0(t)$ of the partition function.
From this level zero character formula, we have extracted the ghosts
multiplicities $N_k$, and using the moments ($\sum_k k^{s+1} N_k$)
of those multiplicities, by employing a novel application of
Pad\'{e} approximants, we were able to compute higher level
character formulas $Z_h(t)$ of pure spinors (up to the twelfth
mass-level $h=12$). We have found that our results are in agreement
with the results found in \cite{PS} (up to the fifth mass-level
$h=5$) where the fixed point technique was used.

We mention a subtle computational issue related to the fixed point
formula \cite{PS,Berkovits:2005hy}. In general for $SO(2d)$ pure
spinors the number of fixed points is $2^{d-1}$ and the complexity
of summing over these fixed points (as it was also noted in
\cite{Berkovits:2005hy}) grows exponentially with $N=2^{d-1}$. On
the other hand the computations shown in this letter are less
complicated, so our technique can be used as an alternative (to the
fixed point technique) easier (computationally) way to get the
character formulas for higher-dimensional pure spinors.

So far, in this work, we have computed the partition function
without the spin dependence on the states. Spin dependence is
crucial if we want to prove that the full partition function
(including the contribution of the worldsheet matter sector)
correctly reproduces the light cone superstring spectrum \cite{PS}.
Therefore, it would be interesting to know the character formula
with the spin dependence in the ghosts-for-ghosts scheme. We left
this issue as a future work.

It would be nice to see whether our technique will also apply to
other constrained systems like strings moving on algebraic surfaces.
Another possible application would be the computation of the
partition function of eleven-dimensional pure spinors, this can be
an interesting issue because an attempt at quantization of the
supermembrane was given by using eleven-dimensional pure spinors
\cite{Berkovits:2002uc}.

\section*{Acknowledgements}
I would like to thank Nathan Berkovits, Yuri Aisaka and Nikita
Nekrasov for useful discussions. I wish also to thank Diany Ruby who
assisted with the proofread of the manuscript. This work is
supported by FAPESP grant 04/09584-6.

\appendix
\setcounter{equation}{0}
\def\thesection{\Alph{section}}
\renewcommand{\theequation}{\Alph{section}.\arabic{equation}}

\section{Computation of higher level character formulas}
Higher level character formulas $Z_h(t)$ can be obtained from the
formal expression (3.12) as follows. Performing a Taylor expansion
of the expression (3.12) around $q=0$, we have
\begin{eqnarray}
Z(q,t)&=&Z_0(t) + \sum_{h=1}^{\infty} \frac{q^h}{h!} {\del^h \over
\del q^h}Z(q,t)\big|_{q=0}
\;\;\;\;\;\;\;\; \\
&=& Z_0(t)\Big[1 +  \sum_{h=1}^{\infty} f_h(t) q^h \Big]\, ,
\nonumber
\end{eqnarray}
where the level $h$ function $f_h(t)$ has been defined as the
expression (4.4). To obtain these level $h$ functions, we are going
to use a method based on Pad\'{e} approximants. Let us explain our
method by computing in detail the level one function $f_1(t)$.

From the expression (4.4), we derive the following expression for
the level one function
\begin{eqnarray}
f_1(t)=\sum_{k=1}^\infty N_k(t^{k}+t^{-k}) \, ,
\end{eqnarray}
expanding the RHS of (A.2) around $t=1$ and keeping terms up to some
order (relevant for the computations to be done next), we get
\begin{eqnarray}
f_1(t)&=&2 \sum_{k=1}^\infty N_k + (t-1)^2 \sum_{k=1}^\infty k^2 N_k
- (t-1)^3 \sum_{k=1}^\infty k^2 N_k + \cdots
\end{eqnarray}
Applying the formula (3.9) to find the even moments $\sum_k N_k$,
$\sum_k k^2 N_k$ and replacing them into the equation (A.3), we
obtain
\begin{eqnarray}
f_1(t)=22 + 4(t - 1)^2 - 4(t - 1)^3
    + \cdots
\end{eqnarray}

Using Pad\'{e} approximants, we express the function $f_1(t)$ as a
rational function
\begin{eqnarray}
f_1(t) \cong f^{[M/N]}_1(t)= \frac{1+\sum_{j=1}^{M} p_j t^j
}{\sum_{j=0}^{N} q_j t^j} \, ,
\end{eqnarray}
for instance, as a pedagogical illustration let us compute
explicitly the $[2/1]$ Pad\'{e} approximant of $f_1(t)$
\begin{eqnarray}
f^{[2/1]}_1(t)= \frac{ 1 + p_1 t + p_2 t^2 }{ q_0 + q_1 t} \, ,
\end{eqnarray}
expanding the RHS of (A.6) around $t=1$, we get
\begin{eqnarray}
f^{[2/1]}_1(t)&=& \frac{1+p_1+p_2}{q_0+q_1}+\frac{p_1 q_0 + 2 p_2
q_0 - q_1 + p_2 q_1}{(q_0 + q_1)^2}(t-1) \\
&+& \frac{p_2 q_0^2 - p_1 q_0 q_1 + q_1^2}{(q_0 +
q_1)^3}(t-1)^2-\frac{p_2  q_1 q_0^2- p_1 q_0 q_1^2 + q_1^3}{(q_0 +
q_1)^4}(t-1)^3 + \cdots \nonumber
\end{eqnarray}

Equating the coefficients of $(t-1)^0$, $(t-1)^1$, $(t-1)^2$,
$(t-1)^3$ in equations (A.4) and (A.7), we get 4 equations for the
unknown coefficients $p_1$, $p_2$, $q_0$, $q_1$
\begin{eqnarray}
\frac{1+p_1+p_2}{q_0+q_1}&=&22 \, , \\
\frac{p_1 q_0 + 2 p_2 q_0 - q_1 + p_2 q_1}{(q_0 + q_1)^2} &=&0 \, ,
\nonumber \\
\frac{p_2 q_0^2 - p_1 q_0 q_1 + q_1^2}{(q_0 + q_1)^3}&=&4 \, ,
\nonumber
\\
\frac{p_2  q_1 q_0^2- p_1 q_0 q_1^2 + q_1^3}{(q_0 + q_1)^4}&=&4 \, ,
\nonumber
\end{eqnarray}
solving these system of equations (A.8) we obtain
\begin{eqnarray}
p_1= \frac{7}{2} \, , \;\;\;\; p_2=1  \, , \;\;\;\; q_0=0  \, ,
\;\;\;\; q_1=\frac{1}{4} \, .
\end{eqnarray}

Computations of higher Pad\'{e} approximants follows in the same way
as it was shown above. The results of these computations are given
in the following table
\begin{align}
\begin{tabular}{|c|c|c|}
    \hline
$[M/N]$  &    $p_1$, $p_2$, $\cdots$, $p_M$   &  $q_0$, $q_1$, $\cdots$, $q_N$       \\
    \hline
[2/1]   & $7/2$, 1  & 0, $1/4$  \\
\hline  [2/2] & $20/23$, 1  & $1/46$, $2/23$, $1/46$  \\
\hline [3/1]   & \,\,\,\,\, 1, $18/25$, $-2/25$ \,\,\,\,\,\,\,\, & $1/50$, $1/10$  \\
\hline [1/3] & $43/23$ & \,\,\,\, $75/5566$, $875/5566$, $-135/2783$, $1/121$  \,\,\,\, \\
\hline [3/2]  & $20/23$, $1$, $0$  & $1/46$, $2/23$, $1/46$  \\
\hline [2/3] & $20/23$, $1$  & $1/46$, $2/23$, $1/46$, $0$   \\
\hline [3/3] & $20/23$, $1$, $0$  & $1/46$, $2/23$, $1/46$, $0$  \\
\hline [4/4]  & $20/23$, $1$, $0$, $0$  & $1/46$, $2/23$, $1/46$, $0$, $0$  \\
\hline
\end{tabular}
\end{align}

As we can see by explicit computations, the Pad\'{e} approximants
are approaching to the rational function $(46 + 40 t + 46 t^2)/(1 +
4 t + t^2)$, and therefore we take this function as being the level
one function $f_1(t)$
\begin{eqnarray}
f_1(t)=\frac{46 + 40 t + 46 t^2}{1 + 4 t + t^2} \, .
\end{eqnarray}
By multiplying this function (A.11) with the level zero character
$Z_0(t)$, we get
\begin{eqnarray}
Z_1(t)=\frac{46 - 144 t + 116 t^2 + 16 t^3 - 16 t^5 - 116 t^6 + 144
t^7 - 46 t^8}{(t-1)^{16}} \, ,
\end{eqnarray}
and therefore, we correctly reproduce the level one character
formula given in \cite{PS,Grassi:2005jz}.

For the next level $h=2$, by using the same strategy shown above, we
have found that the Pad\'{e} approximant computation gives the
following result for the level two function
\begin{eqnarray}
f_2(t)=\frac{-1 + 12 t - 67 t^2 + 248 t^3 + 319 t^4 + 628 t^5 + 319
t^6 + 248 t^7 -
  67 t^8 + 12 t^9 - t^{10}}{t^4(1 + 4 t + t^2)}. \nonumber
\end{eqnarray}
By multiplying this level two function $f_2(t)$ with the level zero
character $Z_0(t)$, we correctly reproduce the level two character
formula found in \cite{PS}.

Computation of higher level functions $f_h(t)$ by means of Pad\'{e}
approximants, suggest us that these functions can be written like
\begin{eqnarray}
f_h(t)= \frac{\sum_{i=0}^{2h+6} C_{i,h} t^i}{t^{h+2}(1 + 4 t + t^2)}
\, .
\end{eqnarray}
We have computed the $C_{i,h}$ coefficients up to the level $h=12$,
the results are given in the tables (4.6) and (4.7) of section 4.
Multiplying the functions $f_h(t)$ with the level zero character
formula $Z_0(t)$, we obtain the characters $Z_h(t)$. We have
compared our first five character formulas with the formulas given
in \cite{PS} and we have found agreement.


\end{document}